\documentclass[fleqn,usenatbib]{mnras}

\usepackage{newtxtext,newtxmath}

\usepackage[T1]{fontenc}

\DeclareRobustCommand{\VAN}[3]{#2}
\let\VANthebibliography\thebibliography
\def\thebibliography{\DeclareRobustCommand{\VAN}[3]{##3}\VANthebibliography}

\usepackage{graphicx}	
\usepackage{amsmath}	
\usepackage{tablefootnote}

\title[TESS Hot Jupiter Occurrence Rates]{Exploring the Dependence of Hot Jupiter Occurrence Rates on Stellar Mass with TESS}

\author[Beleznay \& Kunimoto]{
Maya Beleznay,$^{1}$\thanks{E-mail: mayabel@mit.edu}
M. Kunimoto,$^{1}$
\\
$^{1}$Department of Physics and Kavli Institute for Astrophysics and Space Research, Massachusetts Institute of Technology, Cambridge, MA 02139, USA
}

\date{Accepted 2022 July 23. Received 2022 July 18; in original form 2022 May 13}

\pubyear{2022}

\begin{document}
\label{firstpage}
\pagerange{\pageref{firstpage}--\pageref{lastpage}}
\maketitle

\begin{abstract}

We present estimates for the occurrence rates of hot Jupiters around dwarf stars based on data from the Transiting Exoplanet Survey Satellite (TESS) Prime Mission. We take 97 hot Jupiters orbiting 198,721 AFG dwarf stars (ranging in mass from $0.8M_{\odot}$ to $2.3M_{\odot}$) from an independent search for hot Jupiters using TESS Prime Mission data. We estimate our planet sample's false positive rates as $14\pm7\%$ for A stars, $16\pm6\%$ for F stars, and 0\% for G stars. We find hot Jupiter occurrence rates of $0.29 \pm 0.05\%$ for A stars, $0.36 \pm 0.06\%$ for F stars and $0.55 \pm 0.14\%$ for G stars, with a weighted average across AFG stars of $0.33\pm0.04\%$. Our results show a correlation between higher hot Jupiter abundance and lower stellar mass, and are in good agreement with occurrence rates found by \textit{Kepler}. After correcting for the presence of binaries in the TESS stellar sample, we estimate a single-star hot Jupiter occurrence rate of $0.98\pm0.36\%$ for G stars. This is in agreement with results from radial velocity (RV) surveys, indicating that stellar multiplicity correction is able to resolve the discrepancy between hot Jupiter occurrence rates based on transits and RVs.
\end{abstract}





\section{Introduction}\label{sec:sec1}
Hot Jupiters are giant, gaseous planets that orbit extremely close to their host stars. Their existence has inspired multiple theories of planet formation and migration, such as inward planetary migration and in situ formation via core accretion for orbits close to host stars \cite[e.g.][]{batygin2016, Dawson2018}. 

Hot Jupiters are the easiest planets to detect and characterize due to their large radii and very short orbits, allowing for the observation of multiple and significant transits. However, exoplanet demographics studies have shown that hot Jupiters are relatively rare. Radial velocity (RV) surveys yield occurrence rates of $\sim0.8 - 1.2\%$ \cite[e.g.][]{mayor2011, wright2012, Wittenmyer2020} around sun-like stars. RV surveys have also established a strong positive correlation between host star metallicity and hot Jupiter occurrence \citep[e.g.][]{johnson2010, ghezzi2018}. Studies from the \textit{Kepler} transit survey have found even lower hot Jupiter abundances of $\sim0.4 - 0.6\%$ \citep{howard2012, fressin2013, petigura2018, kunimoto2020}. The evident discrepancy between RV surveys and \textit{Kepler} has been discussed at length with several possible explanations, such as different host star metallicities \cite[e.g.][]{wright2012, hinkel2014, Guo2017}, evolutionary stage \cite[e.g.][]{johnson2010}, stellar multiplicity rate \cite[e.g.][]{wang2014, Moe2021}, and false positive rate \cite[e.g.][]{wang2015}.

Both the \textit{Kepler} mission and RV surveys also hint at stellar mass affecting planetary properties, though there are still open questions regarding the extent to which this is true for hot Jupiters. \textit{Kepler} has been used to determine the occurrence rates primarily of small planets around FGK dwarf stars, and the stellar sample contained very few A stars. While some \textit{Kepler} studies have indicated planet occurrence tends to increase around less massive stars \citep{mulders2015, kunimoto2020}, \textit{Kepler} was unable to comment on the existence of this trend out to A type stars. Meanwhile, RV studies have found giant planet occurrence increases with stellar mass \cite[e.g.][]{Bowler2010, johnson2010, Gaidos2013, ghezzi2018}. Further exploring the discrepancies between RV and transit surveys and the extent to which hot Jupiter occurrence depends on stellar mass could provide information about the conditions under which these planets formed and evolved.

\subsection{The Role of TESS in Hot Jupiter Demographics}\label{sec1.1}

The Transiting Exoplanet Survey Satellite (TESS) was launched on April 18, 2018 aboard a SpaceX Falcon 9 rocket from Cape Canaveral, with the goal of detecting transiting planets around nearby bright stars, especially those that are suitable targets for follow-up \citep{ricker2015}. The satellite measures the brightness of millions of stars landing in $24\times96$ degree ``sectors'' of the sky, each spanning 27.4 days of observations. This observing baseline renders planets with orbital periods $P\lesssim 13$ days ideal targets for detection, as they will have had time to make two full orbits around their host star over the course of a single TESS sector, thus yielding multiple transits. With periods of less than 10 days and large planet radii causing deep transit depths, hot Jupiters are the subgroup of planets most easily detectable by TESS. The all-sky survey approach of TESS also enables the discovery of orders of magnitude more hot Jupiters than have been found by previous surveys.
        
TESS data was previously used to estimate the occurrence rate of hot Jupiters by \citet{zhou2019}, where the estimated occurrence rate of $0.41\pm0.1\%$ for AFG stars brighter than $T_{\text{mag}} = 10$ mag is in good agreement with results found by studies of \textit{Kepler} data. By separating stars into A, F, and G-type bins, they also concluded that there is no dependence of hot Jupiter occurrence rate on stellar mass. However, the estimates from \citet{zhou2019} were performed early in the TESS mission, with planet candidates found only in the first 7 sectors of data included. Given that several years of TESS observations have since been completed, now is the time to revisit the TESS hot Jupiter occurrence rate. In this work, we will use planets and planet candidates identified over the course of the full TESS Prime Mission (26 sectors) to improve hot Jupiter occurrence rate estimates for AFG stars.

\subsection{Paper overview}
In \S \ref{sec2}, we describe the TESS observations used in this project (i.e. the stellar and planet samples), which are chosen similarly to those defined by \citet{zhou2019} to allow close comparisons between our results. In \S \ref{sec3}, we present our analysis, including estimates of false positive rate among TESS hot Jupiters, the process for determining the completeness of our sample, and the final occurrence rate calculation. We present results in \S \ref{sec4}, including discussions of our findings of stellar mass dependence, changes with stellar sample selection, the assumptions used in our false positive calculations, and agreement with \textit{Kepler} and RV. We summarize our findings and discuss the future of TESS in refining exoplanet demographics in \S \ref{sec5}.


\section{Data}\label{sec2}
    \subsection{Stellar sample}\label{sec2.1}
    
        To enable comparison with previous occurrence rates from \textit{Kepler}, we will focus on hot Jupiters around main sequence stars. Following \citet{zhou2019}, we further consider only AFG stars (stellar mass between $0.8$ and $2.3 M_{\odot}$) on the main sequence, and only those that are bright ($T_{\text{mag}} < 10.5$ mag) to make use of planet candidates more readily available to the follow-up community.
        
        Our source of bright main sequence stars is the TESS Input Catalog \citep[TIC;][]{stassun2019}, which contains stellar parameters for targets observed by TESS. A subset of this catalog is the Candidate Target List (CTL), which in v8.01 is comprised of 9.5 million stars identified as the best potential targets for planet detection. The CTL includes properties for all likely dwarf stars with $T_{\text{mag}} < 13$ mag, as well as fainter K and M dwarf stars from the specially curated Cool Dwarf Catalog \citep{muirhead2018}. We downloaded the CTLv8.01 from the Mikulski Archive for Space Telescopes (MAST).\footnote{\url{https://archive.stsci.edu/tess/tic_ctl.html}} We find 392,980 bright AFG stars in the CTL.
        
        Some of these stars may be subgiants, which will have slightly evolved off the main sequence. Using a cut of log$g > 4.1$ as suggested by \citet{stassun2019}, we find 190,604 likely dwarfs among bright AFG stars. However, main sequence stars with different masses will have different typical surface gravities. We alternatively place each bright AFG CTL star on or off the main sequence based on MIST stellar evolution models evaluated with the \texttt{isochrones} Python package \citep{morton2015}. Given a star's stellar mass and assuming solar metallicity, we use \texttt{isochrones} to estimate the surface gravities expected for the Zero-Age Main Sequence (ZAMS) and Terminal-Age Main Sequence (TAMS) evolutionary stages, and determine if the star is on the main sequence if its log$g$ lies between these values.

        Not all of these stars were observed by TESS in its Prime Mission, which had an overall sky coverage of $\sim73\%$. We checked each star for a light curve produced by the Quick-Look Pipeline \cite[QLP;][]{Huang2020}, which extracts light curves for all stars brighter than $T_{\text{mag}} = 13.5$ mag from TESS full-frame images (FFIs)\footnote{\url{https://archive.stsci.edu/hlsp/qlp}}. In total, 198,721 bright AFG main sequence stars identified using our \texttt{isochrones} cut have available light curves, while 141,518 stars identified using the simpler cut on log$g$ have light curves. 221,945 unique stars identified through either cut have QLP light curves from the Prime Mission.
        
        To allow for the assessment of possible hot Jupiter occurrence rate dependence on stellar mass, we create three stellar mass bins using the mass ranges defined by \citet{zhou2019}, corresponding to $0.8 < M_{s} \leq 1.05M_{\odot}$ for G stars 39,747 stars), $1.05 < M_{s} \leq 1.4M_{\odot}$ for F stars (95,561 stars), and $1.4 < M_{s} \leq 2.3M_{\odot}$ for A stars (86,637 stars), as summarized in Table \ref{tab:stars}.
        
        \begin{table}
            \centering
            \begin{tabular}{c|c|c|c}
            \hline
                Type & Using \texttt{isochrones} & Using log$g$ & Using either cut\\
            \hline
                A & 78,179 & 35,962 & 86,637\\
            \hline
                F & 86,289 & 67,688 & 95,561\\
            \hline
                G & 34,253 & 37,868 & 39,747\\
            \hline
                AFG & 198,721 & 141,518 & 221,945\\
            \end{tabular}
            \caption{Breakdown of the subset of bright ($T < 10.5$ mag) AFG main sequence stars from the CTLv8.01 with QLP Prime Mission light curves, depending on whether we identify main sequence stars using the expected log$g$ for dwarf stars as computed with \texttt{isochrones}, or alternatively using a cut of log$g > 4.1$.}
            \label{tab:stars}
        \end{table}
        
   
        
    \subsection{Planet sample}\label{sec2.2}
        We designed an independent search for hot Jupiters using the 221,945 QLP Prime Mission multi-sector light curves. In order to ensure the same pre-search processing elsewhere in our occurrence rate analysis (see \S\ref{sec3.2}), before searching we detrended the raw QLP time series using the biweight time-windowed slider from the \texttt{wotan} detrending package \citep{Hippke2019}, with a window length of 0.5 days. The search was performed with a Box-Least Squares transit search algorithm \cite[BLS;][]{Kovacs2002} as implemented in \texttt{cuvarbase},\footnote{\url{https://github.com/johnh2o2/cuvarbase}} a Python library for fast time-series analysis on CUDA GPUs. We set the period range of our search to be 0.9 to 10 days, which we define as the periods appropriate for hot Jupiters, and set detection criteria of at least two transits and a signal-to-noise ratio (S/N) of at least 10.
        
        We ran tests on the 22,317 signals passing our detection criteria using the pipeline developed by \citet{Kunimoto2022} for automated vetting of planets in TESS FFI light curves. These tests assess the strength of planet candidacy according to standard diagnostics, such as the uniqueness of the event in the light curve, a lack of significant odd versus even transit depth differences or secondary events in the light curves that could suggest an eclipsing binary false positive, or no significant centroid offsets that could indicate the transit signal is from a nearby or background star. We then manually inspected surviving candidates to identify hot Jupiters. In order to minimize bias in this process, we did not refer to the list of known TESS Objects of Interest (TOIs) already detected for these stars.
        
        Finally, we fit transit models to each light curve to derive a uniform list of each candidate's properties. The transit models were parameterized by orbital period ($P$), time of first transit time ($t_{0}$), planet-to-star radius ratio ($R_{p}/R_{s}$), impact parameter ($b$), mean flux offset ($f_{0}$), and stellar density ($\rho_{\odot}$). We assumed circular orbits, and set quadratic limb-darkening parameters equal to their interpolated values based on \citet{claret2017}. We explored the parameter space with the \texttt{exoplanet} Python package \citep{exoplanet:joss}, with 4 chains sampled for 3000 tuning steps and 3000 draw steps each.
        
        We use the same definition as \citet{zhou2019} to define hot Jupiters as those with radii $0.8 < R_p < 2.5R_J$ ($9 < R_p < 28 R_{\oplus}$), orbital periods $0.9 < P < 10$ days, and transit impact parameters $b < 0.9$. While hot Jupiter TOIs with $b > 0.9$ exist, the grazing transit geometry transits makes determining the correct planet radius complicated, and they are harder to distinguish from grazing eclipsing binary stars. After filtering candidates matching these properties according to our transit model fit results, we found 100 hot Jupiters around bright AFG main sequence stars from the TESS Prime Mission (97 around our isochrones stellar sample; 68 around our log$g$ stellar sample).

        \subsubsection{Comparison to the TOI Catalog}
        
        Our Prime Mission TESS hot Jupiter catalog bears many resemblances to the hot Jupiters in the TOI Catalog. TOIs are planet candidates that were initially detected through various planet search pipelines, primarily NASA's Science Processing Operations Center \citep[SPOC;][]{Jenkins2016} and QLP, and then passed manual inspection by the TESS Science Office (TSO) in a process similar to our manual vetting stage. TOIs may later receive follow-up by the TESS Follow-Up Observing Program Working Group (TFOPWG) to determine whether they are real planets or false positives. The TFOPWG assigns dispositions such as known planet (KP; planets already discovered/confirmed by previous surveys), confirmed planet (CP; planets discovered/confirmed by TESS), false positive (FP; namely nearby or bound eclipsing binary stars which commonly mimic planet signals), ambiguous planet candidate (APC; highly likely but not confirmed false positive), and planet candidate (PC; no disposition).
        
        We accessed the TOI Catalog from the Exoplanet Follow-Up Observation Program website\footnote{\url{https://exofop.ipac.caltech.edu/tess/}} \citep{ExoFOP} on 2022 June 23, and refer to Prime Mission TOIs as those that are also members of the Prime Mission TOI Catalog \citep{guerrero2021}. Our search pipeline recovered all Prime Mission hot Jupiter TOIs that are associated with our stellar samples. However, our manual inspection stage failed 14 of these TOIs, and thus they did not make it into our final catalog. We labeled 4 of these TOIs (TOI-184.01, 585.01, 1455.01, 1645.01) as FPs due to the presence of significant secondary eclipses consistent with an eclipsing binary. Each of these TOIs have current dispositions of either APC or FP, supporting this interpretation. We labeled another 9 TOIs (TOI-387.01, 610.01, 951.01, 1110.01, 1120.01, 1335.01, 1390.01, 1556.01, 1888.01) as FPs due to significant offsets visible in the difference image data products produced by our vetting pipeline, indicating the transit sources are off-target. According to observing notes on ExoFOP, all of these TOIs are considered off-target EBs. We labeled only one PC TOI (TOI-1521.01) an FP, though our centroid analysis strongly indicates that this is an off-target signal originating from TIC-366074071.
        
        We additionally found 6 TOIs that are not members of the Prime Mission TOI Catalog: one KP (TOI-4486.01) and 5 PCs identified over the course of the Extended Mission (TOI-2336.01, 2360.01, 4492.01, 4497.01, 4768.01). We believe that the optimization of our pipeline for searching only hot Jupiter periods is what enabled the detection of these signals with Prime Mission data alone, while they were missed by other pipelines.
        
        Finally, we detected one promising hot Jupiter candidate that is not yet a TOI but passed our vetting. As mentioned earlier, we believe that TOI-1521.01 is an off-target FP originating from TIC-366074071, which is also member of our stellar sample. We found a 4.2-day signal consistent with a hot Jupiter with radius $R_{p} = 16.3 R_{\oplus}$ around this star. For the purposes of our analysis, we included it in our planet catalog and assigned it a PC disposition.
        
        Overall, our hot Jupiter catalog consists of 99 TOIs and one new planet candidate. 54 are already known KPs, 10 have since been confirmed by TESS,\footnote{TOI-135.01 \citep{Jones2019}; TOI-624.01 \citep{zhou2019}; TOI-625.01 \citep{zhou2019}; TOI-626.01 \citep{martinez2020}; TOI-628.01 \citep{rodriguez2021}; TOI-640.01 \citep{rodriguez2021}; TOI-1268.01 \citep{subjak2022}; TOI-1333.01 \citep{rodriguez2021}; TOI-1494.01 \citep{schanche2020}; TOI-1516.01 \citep{Kabath2022}} and 7 have since been retired as FPs. The remaining 29 have not yet received final dispositions, and are either PCs or APCs. Our hot Jupiters with these dispositions is shown in Table \ref{tab:tois}, along with planet physical and orbital properties from our transit model fits and host star parameters from the CTL.
        
        \color{black}
        
\begin{table*}
    \centering\setlength{\tabcolsep}{4pt}

    \begin{tabular}{ccccccccccc}
\hline\hline
TOI & TIC & $R_{p} (R_{\oplus}$) & $P$ (days) & $t_{0}$ & $b$ & $R_{s} (R_{\odot}$) & $M_{s} (M_{\odot})$ & log$g$ & Disposition & Sample \\
& & & & (BJD - 2457000) & & & & (log cm s$^{-2}$) & & \\
\hline
- & 366074071 & $16.25_{-0.10}^{+0.10}$ & $                                                      4.181295_{-0.000017}^{+0.000017}$ & $1768.0874_{-0.0006}^{+0.0006}$ & $0.584_{-0.031}^{+0.029}$ & $1.76\pm0.08$ & $1.24\pm0.20$ & $4.043\pm0.087$ & PC & iso \\
102.01 & 149603524 & $15.50_{-0.02}^{+0.02}$ & $                                                      4.411937_{-0.000002}^{+0.000002}$ & $1326.0788_{-0.0001}^{+0.0001}$ & $0.628_{-0.004}^{+0.004}$ & $1.21\pm0.05$ & $1.28\pm0.19$ & $4.377\pm0.080$ & KP & both \\
105.01 & 144065872 & $14.52_{-0.05}^{+0.05}$ & $                                                      2.184627_{-0.000029}^{+0.000028}$ & $1326.5061_{-0.0002}^{+0.0002}$ & $0.743_{-0.007}^{+0.007}$ & $1.24\pm0.06$ & $1.03\pm0.13$ & $4.265\pm0.076$ & KP & both \\
106.01 & 38846515 & $16.76_{-0.04}^{+0.03}$ & $                                                      2.849382_{-0.000003}^{+0.000003}$ & $1326.7450_{-0.0002}^{+0.0002}$ & $0.743_{-0.005}^{+0.005}$ & $1.77\pm0.08$ & $1.44\pm0.24$ & $4.098\pm0.092$ & KP & iso \\
107.01 & 92352620 & $20.70_{-0.08}^{+0.08}$ & $                                                      3.950270_{-0.000142}^{+0.000140}$ & $1328.2990_{-0.0004}^{+0.0004}$ & $0.593_{-0.018}^{+0.017}$ & $1.71\pm0.07$ & $1.14\pm0.15$ & $4.029\pm0.076$ & KP & iso \\
135.01 & 267263253 & $17.20_{-0.05}^{+0.05}$ & $                                                      4.126986_{-0.000078}^{+0.000075}$ & $1325.7837_{-0.0003}^{+0.0003}$ & $0.611_{-0.015}^{+0.014}$ & $1.59\pm0.06$ & $1.46\pm0.25$ & $4.201\pm0.084$ & CP & both \\
143.01 & 25375553 & $17.98_{-0.09}^{+0.09}$ & $                                                      2.311050_{-0.000068}^{+0.000068}$ & $1325.5819_{-0.0004}^{+0.0004}$ & $0.767_{-0.010}^{+0.010}$ & $1.95\pm0.09$ & $1.34\pm0.22$ & $3.982\pm0.092$ & KP & iso \\
185.01 & 100100827 & $15.63_{-0.04}^{+0.04}$ & $                                                      0.941453_{-0.000004}^{+0.000004}$ & $1354.4579_{-0.0001}^{+0.0001}$ & $0.809_{-0.003}^{+0.003}$ & $1.35\pm0.07$ & $1.20\pm0.17$ & $4.259\pm0.087$ & KP & both \\
190.01 & 166739520 & $12.36_{-0.06}^{+0.06}$ & $                                                     10.020993_{-0.000171}^{+0.000174}$ & $1357.5710_{-0.0005}^{+0.0005}$ & $0.581_{-0.023}^{+0.022}$ & $1.28\pm0.06$ & $1.10\pm0.14$ & $4.268\pm0.074$ & KP & both \\
191.01 & 183532609 & $12.84_{-0.06}^{+0.06}$ & $                                                      8.158899_{-0.000218}^{+0.000217}$ & $1358.9200_{-0.0003}^{+0.0003}$ & $0.694_{-0.013}^{+0.012}$ & $1.00\pm0.06$ & $0.99\pm0.12$ & $4.437\pm0.085$ & KP & both \\
... & ... & ... & ... & ... & ... & ... & ... & ... & ... & ...\\
    \end{tabular}
    \caption{The fitted properties for all 100 hot Jupiters uncovered for this occurrence rate study, alongside host star properties from the CTLv8.01 \citep{stassun2019}. The TESS Follow-Up Working Group (TFOPWG) Dispositions were obtained from the Exoplanet Follow-Up Observation Program website \citep{ExoFOP} on 2022 June 23. TIC-366074071.01 is not a known TOI, but for the purposes of our analysis we assigned it a disposition of PC. The Sample column refers to whether the hot Jupiter is associated with our isochrones-selected stellar sample, the log$g$-selected stellar sample, or both. Only the first 10 hot Jupiters are shown here. The full table is available in machine-readable format.}
    \label{tab:tois}
\end{table*}


\section{Analysis}\label{sec3}
\begin{figure}\label{afg_subfig}
\centering
       \includegraphics[width=\linewidth]{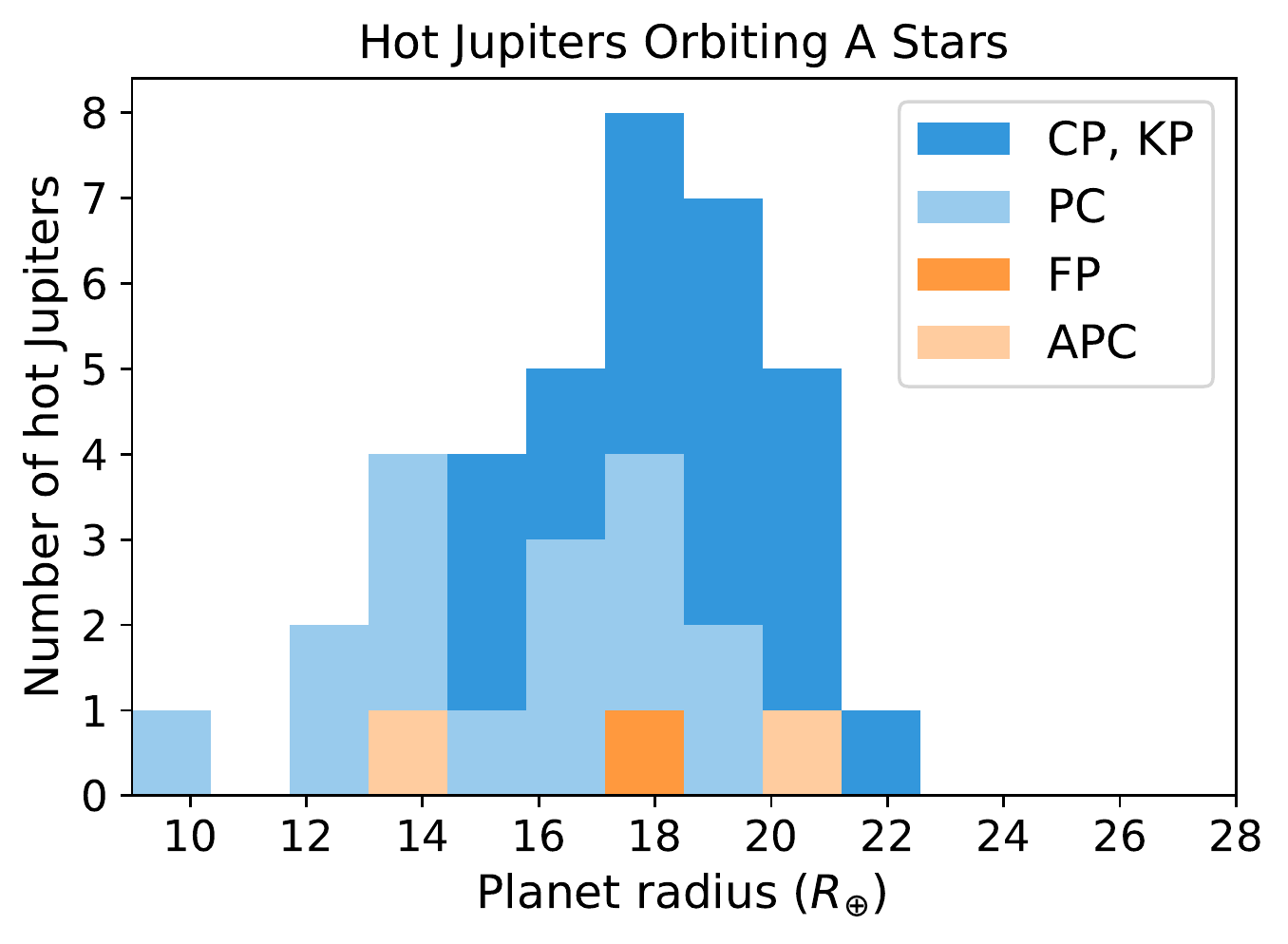}
       \includegraphics[width=\linewidth]{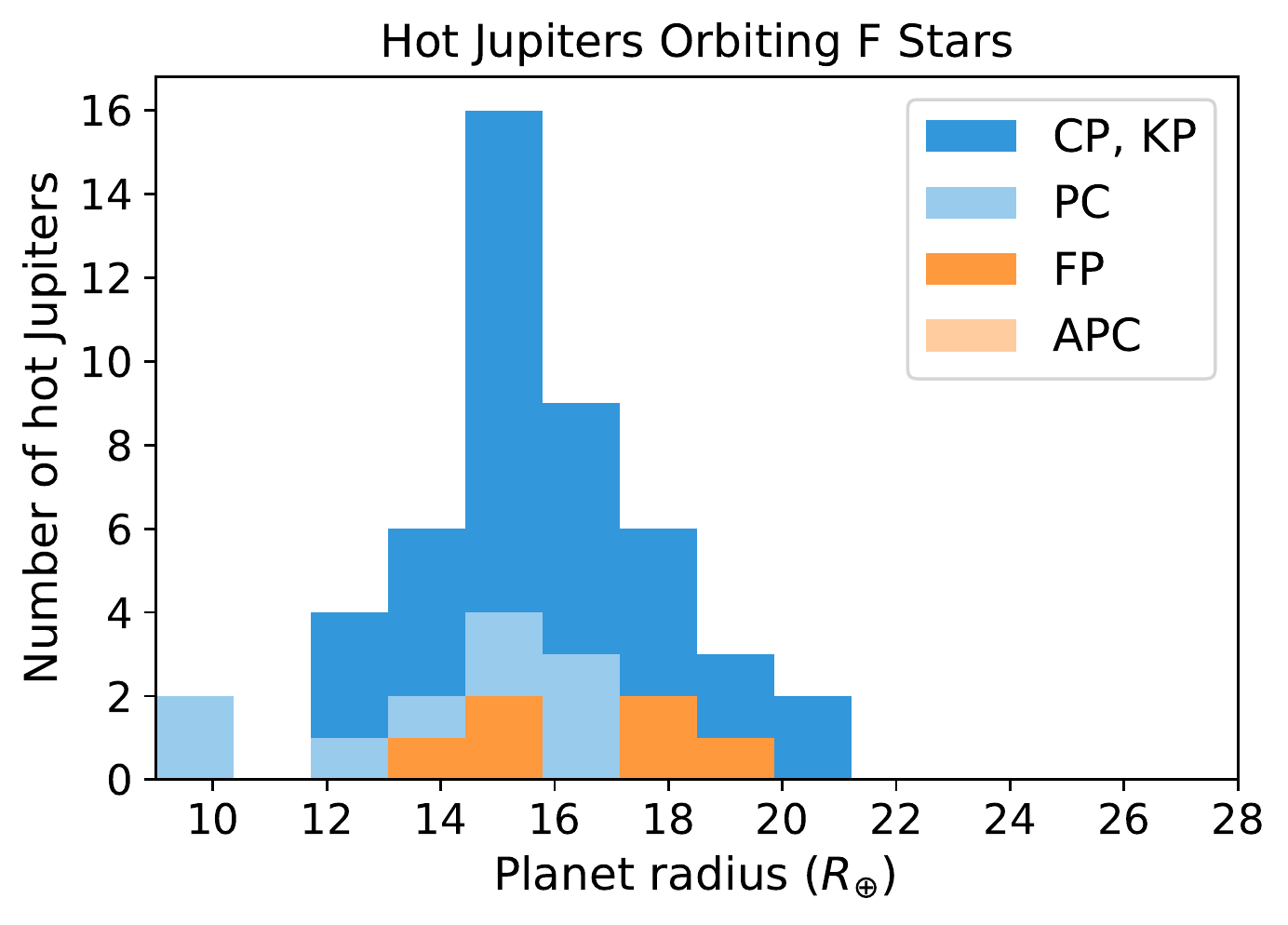}
       \includegraphics[width=\linewidth]{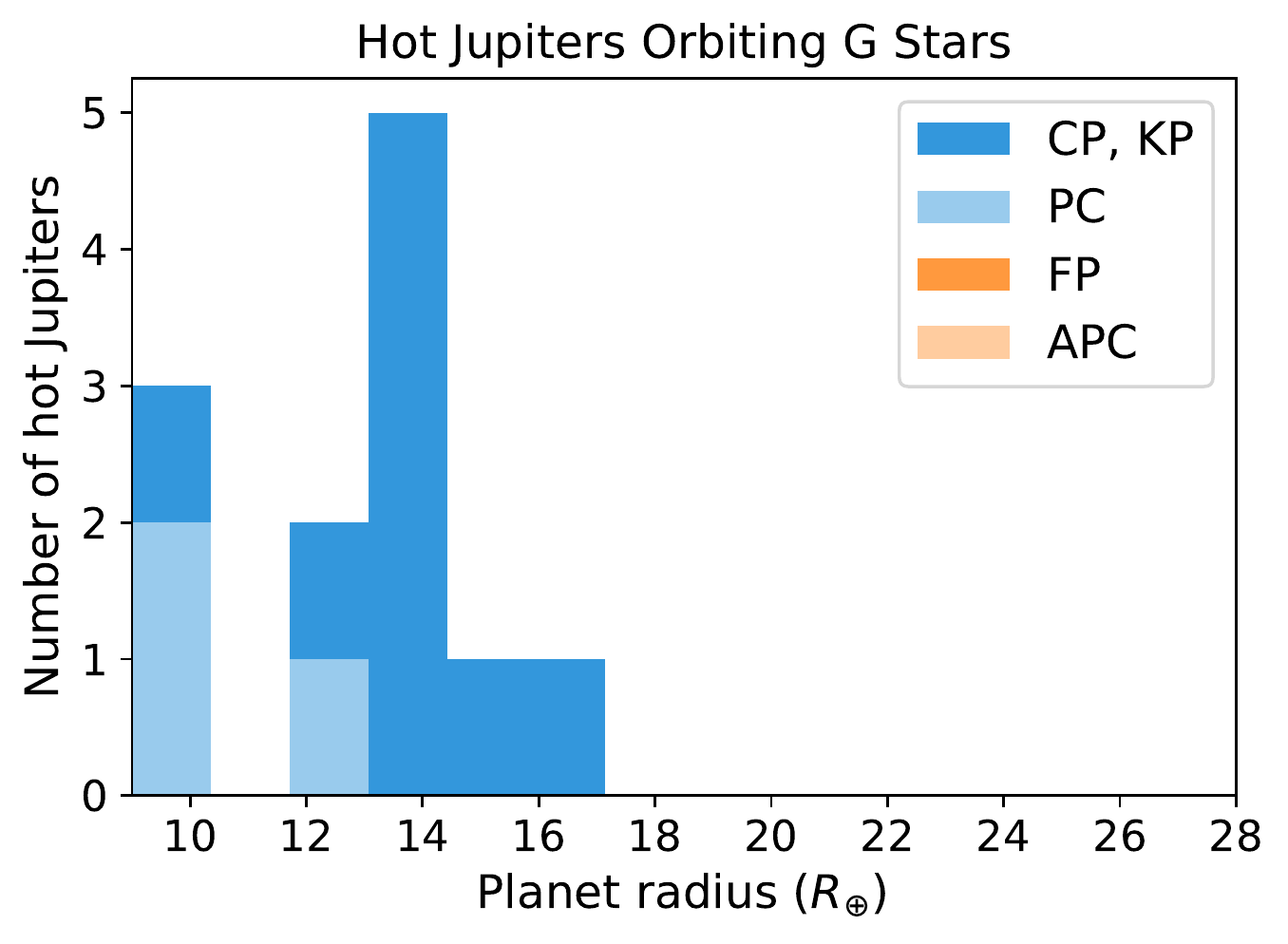}
\caption{The TESS hot Jupiter population for A, F, and G dwarf stars. There are more smaller planets around G stars, while the larger planets are around A stars. Hot Jupiters are color-coded by TESS Follow-Up Observing Program Working Group disposition, showing confirmed and known planets (CPs, KPs), false positives (FPs), ambiguous planet candidates (APCs), and undispositioned planet candidates (PCs).}
\end{figure}

We will use our isochrones-selected stellar sample to perform the baseline analysis discussed in this section. The 97 hot Jupiters and their properties are broken down by stellar type in Figure \ref{afg_subfig}. For reference \citet{zhou2019} found occurrence rates based on 18 hot Jupiters orbiting 47,126 stars. We anticipate finding occurrence rates with lower uncertainty than \citet{zhou2019} due to the much higher number of hot Jupiters and stars in our samples.

    \subsection{False positive rate}\label{sec3.1}
    
    Not all of the hot Jupiters in our catalog are necessarily planets. Among our isochrones-selected sample we find 7 TOIs are FPs compared to 9 CPs and 52 KPs. The remaining  29 have not yet received final dispositions, and are either PCs or APCs.
    
    Information from \textit{Gaia} Data Release 3 \cite[DR3;][]{Gaia2016, Gaia2022} can provide additional information to identify likely false positives not already retired through TFOPWG follow-up. We would treat such hot Jupiters as APCs for the purpose of our analysis. We cross-matched our hot Jupiter hosts with Gaia DR3 and inspected each star's renormalized unit weight error (RUWE) metric. The RUWE describes  the astrometric fit using  the square root of the normalized chi-square. Physically, it describes the wobble of the star's photocenter. RUWE values of $\sim1$ are the solutions to single star systems, while RUWE exceeding 1.4 is an indicator of a possible astrometric companion \citep{khandelwal2022}. Only two of our hot Jupiters hosts have RUWE $> 1.4$ (TOI-630 and TOI-2047), and these TOIs have already been classified as an APC and FP.
    
    \textit{Gaia} DR3 also includes the new Non-Single Star (NSS) flag, which indicates if the star has been identified as an astrometric binary, spectroscopic binary, or eclipsing binary \citep{Gaia2022b}. All of our un-dispositioned hot Jupiters have NSS = 0, meaning a non-single star solution was not found. In summary, we do not find sufficient evidence to reclassify any of our PC hot Jupiters using \textit{Gaia} data.

    When calculating the hot Jupiter occurrence rate, we must correct for the false positive rate among the undispositioned planet candidates. We interpret the false positive rate to be the fraction of hot Jupiters that are FPs out of all hot Jupiters that have a final disposition. We first perform estimates of the false positive rate by treating APCs as FPs, which is the most likely interpretation. We also later perform separate estimates by treating APCs as PCs to assess the impact of this assumption (see \S \ref{secAPC}).
    
    Treating FPs (and APCs) as objects with a false positive probability of 1 and CPs and KPs as those with a false positive probability of 0, the overall false positive rate (FPR) among undispositioned PCs is:
    \par

    \begin{equation}\label{eqn:fpr}
    \text{FPR} = \frac{N_{\text{FP}} + N_{\text{APC}}}{N_{\text{FP}}+N_{\text{APC}} + N_{\text{CP}} + N_{\text{KP}}}.
    \end{equation}
    
    We calculate the false positive rate separately for each stellar mass bin, as shown in Table \ref{tab1}, with uncertainties estimated through the Beta probability distribution.
    
    \begin{table}
        \centering
        \begin{tabular}{|c|c|c|c|c|}
        \hline
         Type & FP, APC & CP, KP & PC & FPR \\ 
         \hline
         A & 1, 2 & 5, 14 & 15 &  $14\pm7\%$\\ 
         \hline
         F & 6, 0 & 3, 29 & 9 & $16\pm6\%$\\
         \hline
         G & 0, 0 & 1, 9 & 3 & $0\%$ \\
         \hline
        \end{tabular}
        \caption{Breakdown of TOIs and their dispositions for each stellar mass bin, alongside false positive rates (FPRs) estimated with Eqn. \ref{eqn:fpr}.}
        \label{tab1}
    \end{table}
    

    \subsection{Completeness}\label{sec3.2}
    
    The completeness of our planet sample must be accounted for prior to calculating the occurrence rate, given that our planet sample represents only a small subset of all hot Jupiters orbiting AFG stars. First, the transit method is insensitive to systems oriented in such a way that their planets do not pass directly in front of their host star, as observed by TESS. To correct for this, we must understand the likelihood of a planet transiting its star. The transit probability is given as
    \begin{equation}\label{eq2}
        \mathcal{P}_{tr} = 0.9\frac{R_s + R_p}{a}
    \end{equation}
    for a star of radius $R_s$, with a planet of radius $R_p$ orbiting at a semimajor axis, $a$ from Kepler's Third Law,
    \begin{equation}\label{eq3}
        a = \bigg(\frac{GM_sP^2}{4\pi^2}\bigg)^{1/3}.
    \end{equation}
    The additional factor of 0.9 in Eqn. \ref{eq2} arises from the fact that we consider planets with impact parameters less than 0.9 to ensure proper detection. We compute each star's average hot Jupiter transit probability by randomly drawing 100 planets around each star with uniformly drawn orbital periods ($0.9 < P < 10$ days) and planet radii ($9 < R_{p} < 28 R_{\oplus}$) and taking the average $\mathcal{P}_{tr}$ computed with Eqn \ref{eq2}.
    
    Incompleteness can also result from imperfect detection probability. For example, our search pipeline may have missed a planet if its S/N was not strong enough, or it did not have at least two transits in the data. A common method for characterizing a pipeline's detection probability is through transit injection and recovery tests \cite[e.g.][]{Christiansen2020}, where the fraction of injected transits that are recovered can empirically describe detection probability over transit properties such as S/N and orbital period. 
    
    We performed injection/recovery tests by randomly generating orbits for the previously drawn 100 hot Jupiters around each star, with uniformly drawn impact parameters ($0 < b < 0.9$) and orbital phases (between 0 and 1), resulting in 22,194,500 total simulated planets. We injected transit models for these planets computed with the \texttt{batman} exoplanet transit modelling package \citep{Kreidberg2015} into the raw QLP light curves, and detrended the light curve with the same procedure that was used for our original planet search. We also subtracted the transit models of known hot Jupiters from relevant light curves before searching to ensure they would not affect recovery. We then used our search pipeline to attempt to detect the injected hot Jupiters with the same detection criteria as our original planet search, and used the ephemeris-matching procedure described by \citet{Mullally2015} to determine whether or not we successfully found the injected signal. Finally, we ran the automated vetting pipeline on all ephemeris-matched detections. We consider an injected hot Jupiter recovered if it was both detected by our search pipeline and passed automated vetting.
    
    While automated vetting had been followed by manual vetting in our original search to identify our hot Jupiter sample, we opted not to include this step given it would require inspection of $\sim20,000,000$ signals. We expect manual inspection would have failed a negligible number of these signals, given hot Jupiters are especially high-S/N and unlikely to be confused for noise and systematics.
    
    The hot Jupiter detection probability for each star, $\mathcal{P}_{det}$, is the fraction of injected planets that were recovered. We find detection probabilities of $\langle\mathcal{P}_{det}\rangle =$ 0.819, 0.966 and 0.978 when averaged over all A, F and G stars, respectively. The overall probability of finding a hot Jupiter around a specific star is $\mathcal{P}_{total} = \mathcal{P}_{det}\mathcal{P}_{tr}$. 
    

    \subsection{Occurrence rate calculation}\label{sec3.3}
    We calculate the occurrence rate, \textit{f}, using a Beta distribution model, following the method of \cite{zhou2019}. The occurrence rate is given by
    \begin{equation}\label{eq10}
        \mathcal{P}(f|n_{obs}, n_{trial}) \sim \text{Beta}(n_{obs},n_{trial}-n_{obs})
    \end{equation}
    where $n_{obs}$ is the number of planets observed in a mass bin and $n_{trial}$ is the effective number of times TESS tried to detect those planets. $n_{obs}$ is a weighted sum given by
    \begin{equation}\label{eq11}
        n_{obs} = \sum_{i = 1}^{n_p}(1-\text{FPR}_i)\omega_i,
    \end{equation}    
    where $\text{FPR}_i$ is the false positive rate for a given mass bin, and $\omega_{i}$ is a weight that depends on the probability that the host star of interest falls within that mass bin. We determine $\omega_{i}$ by assuming a Gaussian probability distribution centred on the mass of the host star, with standard deviation given by the uncertainty in the mass.
    
    We estimate $n_{trial}$ for $N_s$ stars using
     \begin{equation}\label{eq12}
        n_{trial} = \sum_{j = 1}^{N_s}\mathcal{P}_{total,j}
    \end{equation}      
    where $\mathcal{P}_{total,j}$ is the average detection probability for the $j$-th star as calculated in \S \ref{sec3.2}. Given $n_{obs}$ and $n_{trial}$, we calculate the hot Jupiter occurrence rate and its uncertainty using our Beta distribution model in Equation \ref{eq10}.
    

\section{Results and Discussion}\label{sec4}

\begin{figure}\label{fig5}
            \centering
            \includegraphics[width=\linewidth]{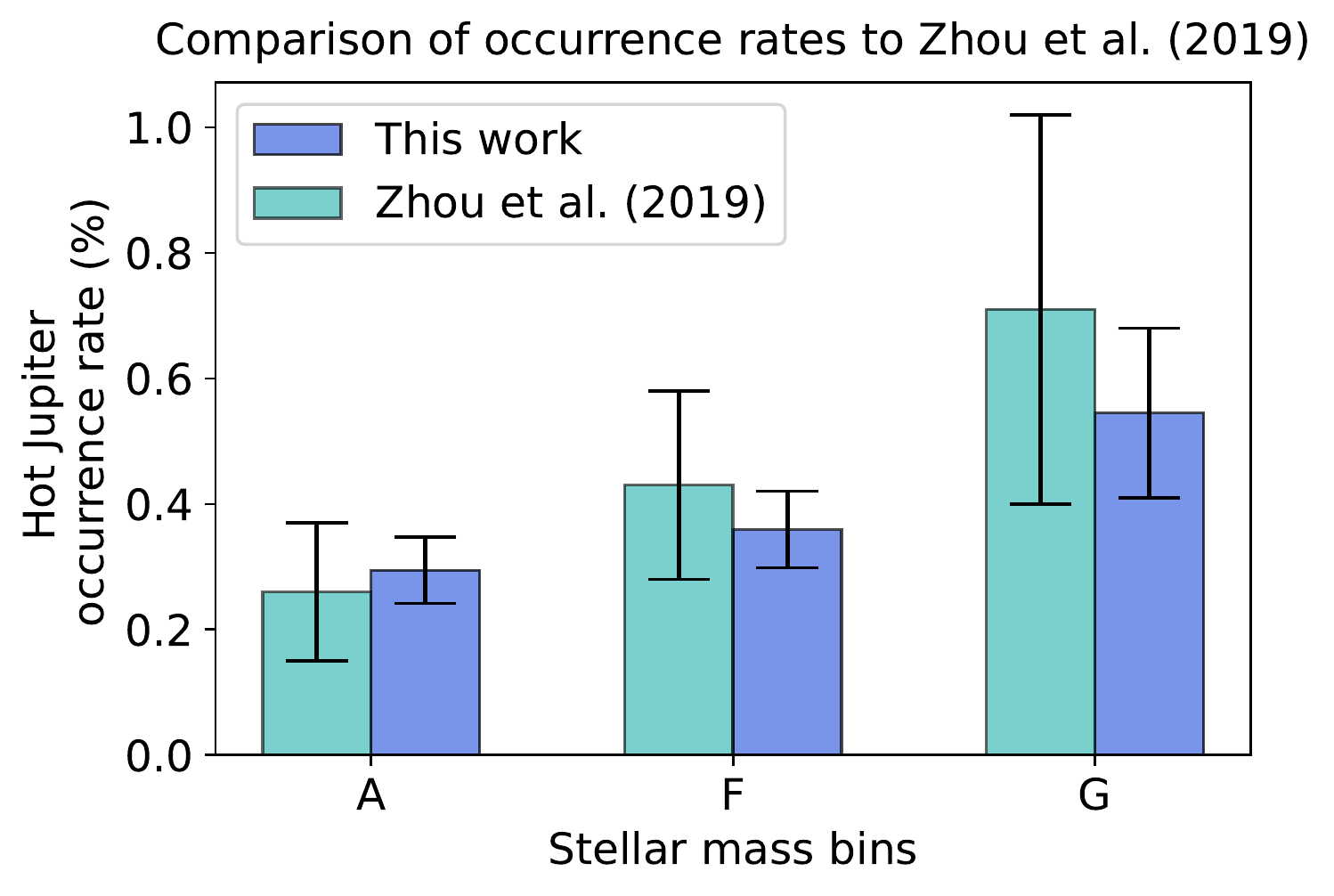}
            \caption{Hot Jupiter occurrence rates for A, F, and G dwarf stars found by this work compared to those by \citet{zhou2019}. The uncertainties on our occurrence rates are roughly half of those from \citet{zhou2019}. Our work makes use of the full 26 sectors from the TESS Prime Mission, while \citet{zhou2019} used only the first 7 sectors. We had access to a much larger sample of hot Jupiters (97 TOIs vs. 18 TOIs) and observed stars (192,250 stars vs. 47,126 stars), which improved the statistical power of the TESS observations.}
        \end{figure}

    We find occurrence rates of  $0.29 \pm 0.05\%$ for A type stars, $0.36 \pm 0.06\%$ for F type stars and $0.55 \pm 0.14\%$ for G type stars. Our results indicate a dependence of hot Jupiter occurrence rate on stellar mass, where hot Jupiter abundances decrease as stellar mass increases. The weighted mean hot Jupiter occurrence rate for AFG stars is $0.33 \pm 0.09\%$. By comparison, \citet{zhou2019} found occurrence rates of $0.26\pm0.11\%$, $0.43\pm0.15\%$, and $0.71\pm0.31\%$. Our results  cur occurrence rate uncertainties by more than half across all three stellar types, yielding the most accurate TESS hot Jupiter occurrence rate calculations for these stellar types to date (Fig. \ref{fig5}). Our full list of the occurrence rates discussed in this section are summarized in Table \ref{tab:results} alongside those from other works.
    
    \begin{table*}
        \centering
        \begin{tabular}{cccccc}
        \hline\hline
            A (\%) & F (\%) & G (\%) & Reference & Detections & Hot Jupiter Definition\\
        \hline
            $0.29\pm0.05$ & $0.36\pm0.06$ & $0.55\pm0.14$ & Baseline results: isochrones sample, APCs treated as FPs & Transit & $9 - 28 R_{\oplus}$, $0.8 - 10$ days\\
            $0.52\pm0.12$ & $0.36\pm0.43$ & $0.43\pm0.12$ & \S\ref{sec4.2}: log$g$ sample, APCs treated as FPs & Transit & $9 - 28 R_{\oplus}$, $0.8 - 10$ days\\
            $0.32\pm0.06$ & $0.36\pm0.06$ & $0.55\pm0.14$ & \S\ref{secAPC}: isochrones sample, APCs treated as PCs & Transit & $9 - 28 R_{\oplus}$, $0.8 - 10$ days\\
            $0.62\pm0.32$ & - & $0.98\pm0.36$ & \S\ref{secRV}: baseline corrected for stellar multiplicity & Transit & $9 - 28 R_{\oplus}$, $0.8 - 10$ days\\
            \hline
            - & $0.21\pm0.10$ & $0.52\pm0.18$ & \citet{kunimoto2020} & Transit & $8 - 16 R_{\oplus}$, $0.78 - 12.5$ days \\
            $0.26\pm0.11$ & $0.43\pm0.15$ & $0.71\pm0.31$ & \citet{zhou2019} & Transit & $9 - 28 R_{\oplus}$, $0.8 - 10$ days\\
            - & - & $0.57_{-0.12}^{+0.14}$ & \citet{petigura2018} & Transit & $8 - 24 R_{\oplus}$, $1 - 10$ days\\
            - & - & $0.45\pm0.10$ & \citet{fressin2013} & Transit & $6 - 22 R_{\oplus}$, $0.8 - 10$ days \\
            - & - & $0.40\pm0.10$ & \citet{howard2012} & Transit & $8 - 32 R_{\oplus}$, $< 10$ days\\
            \hline
            - & - & $0.84_{-0.20}^{+0.70}$ & \citet{Wittenmyer2020} & RV & $M_{p}\sin{i} > 0.3 M_{J}$, $1 - 10$ days\\
            - & - & $1.2\pm0.38$ & \citet{wright2012} & RV & $M_{p}\sin{i} > 0.1 M_{J}$, $< 0.1$ AU\\
            - & - & $0.89\pm0.36$ & \citet{mayor2011} & RV & $M_{p}\sin{i} > 50 M_{\oplus}$, $< 11$ days \\ 
        \end{tabular}
        \caption{Summary of hot Jupiter occurrence rates from this work and others \citep{mayor2011, howard2012, wright2012, fressin2013, petigura2018, kunimoto2020, zhou2019, Wittenmyer2020}. The reported values from \citet{howard2012, petigura2018} are relevant for GK dwarfs, while the reported values from \citet{mayor2011, wright2012, fressin2013, Wittenmyer2020} are relevant for FGK dwarfs or otherwise Sun-like stars. We consider these most comparable to our G star occurrence rates.}
        \label{tab:results}
    \end{table*}

    \subsection{Importance of stellar sample selection}\label{sec4.2}
    
    In \S \ref{sec2.1}, we had chosen our main sequence stellar sample based on isochrones, as informed by each star's mass and log$g$. Alternatively, we could have used a more traditional, simple log$g$ cut on all stars, which has been used by several previous \textit{Kepler} papers \cite[e.g.][]{fressin2013, petigura2018}. To assess the sensitivity of our results on the stellar sample selection, we re-calculated each occurrence rate after having used log$g > 4.1$ to place stars on the main sequence.
    
\begin{figure} \label{fig8}
        \centering
        \includegraphics[width=\linewidth]{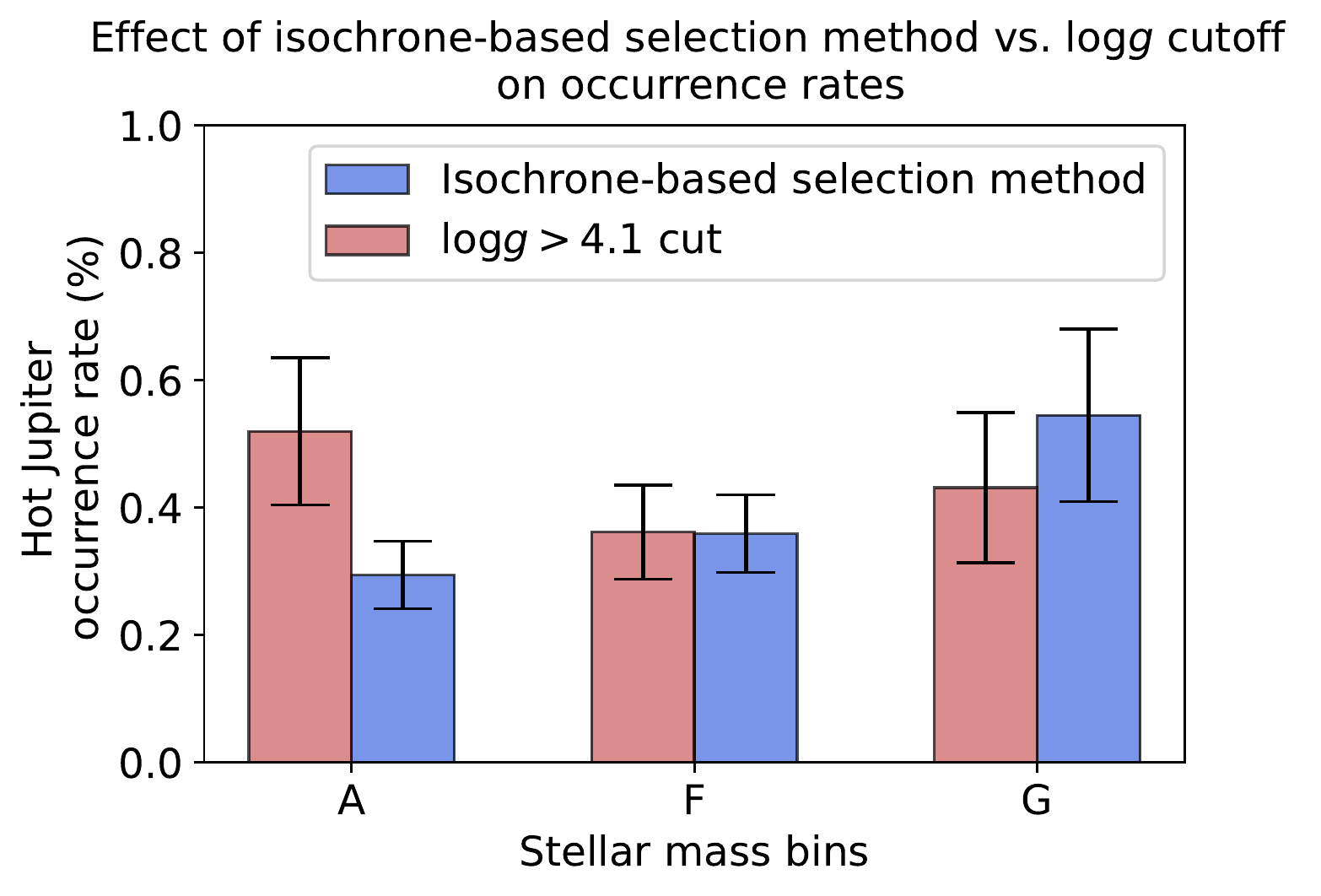}
        \caption{Hot Jupiter occurrence rates for A, F, and G dwarf stars found having used our default isochrone-based selection method to determine the stellar sample, compared to a simpler log$g$ cut. The most significant change is for A stars, for which the log$g$ cut caused the stellar sample size to drop by a factor of two and the resulting occurrence rate to double.}
    \end{figure}

    In  Fig. \ref{fig8} we highlight the different results yielded by stellar samples selected via the log$g$ method and those selected via isochrones. The main stellar group affected was the A-type stars, because the log$g$ cuts are highly biased toward cooler, lower mass stars, thus cutting out a significant portion of the A type stellar sample. We find that our A-type stellar sample drops from 78,179 stars to 35,962 stars, and the associated occurrence rate doubles from $0.26\pm0.05\%$ to $0.52\pm0.12$.

    We find that the number of F stars also decreases when using the log$g$ cut, from 86,289 to 67,688 stars, though the associated occurrence rate remains roughly the same ($0.36\pm0.07\%$). For G type stars, the number of stars increases slightly from 34,253 to  37,868, and the occurrence rate drops to $0.43\pm0.12\%$. The F and G occurrence rates are well within $1\sigma$ of each other (Fig. \ref{fig8}), and we are no longer able to discern a significant trend with stellar mass when using log$g$. We trust our results from isochrone selection over log$g$ selection, given it is a more physically-motivated choice.
    \subsection{Viability of assumptions for APCs}\label{secAPC}
        In \S\ref{sec3.1}, we described the method behind determining the false positive rate, which is later used to characterize $n_{obs}$. Because APCs are highly likely to be FPs, we treated them as FPs when computed the false positive rate. To test the consequences of this assumption, we repeated our analysis with APCs treated as PCs.
        
\begin{figure} \label{fig6}
        \centering
        \includegraphics[width=\linewidth]{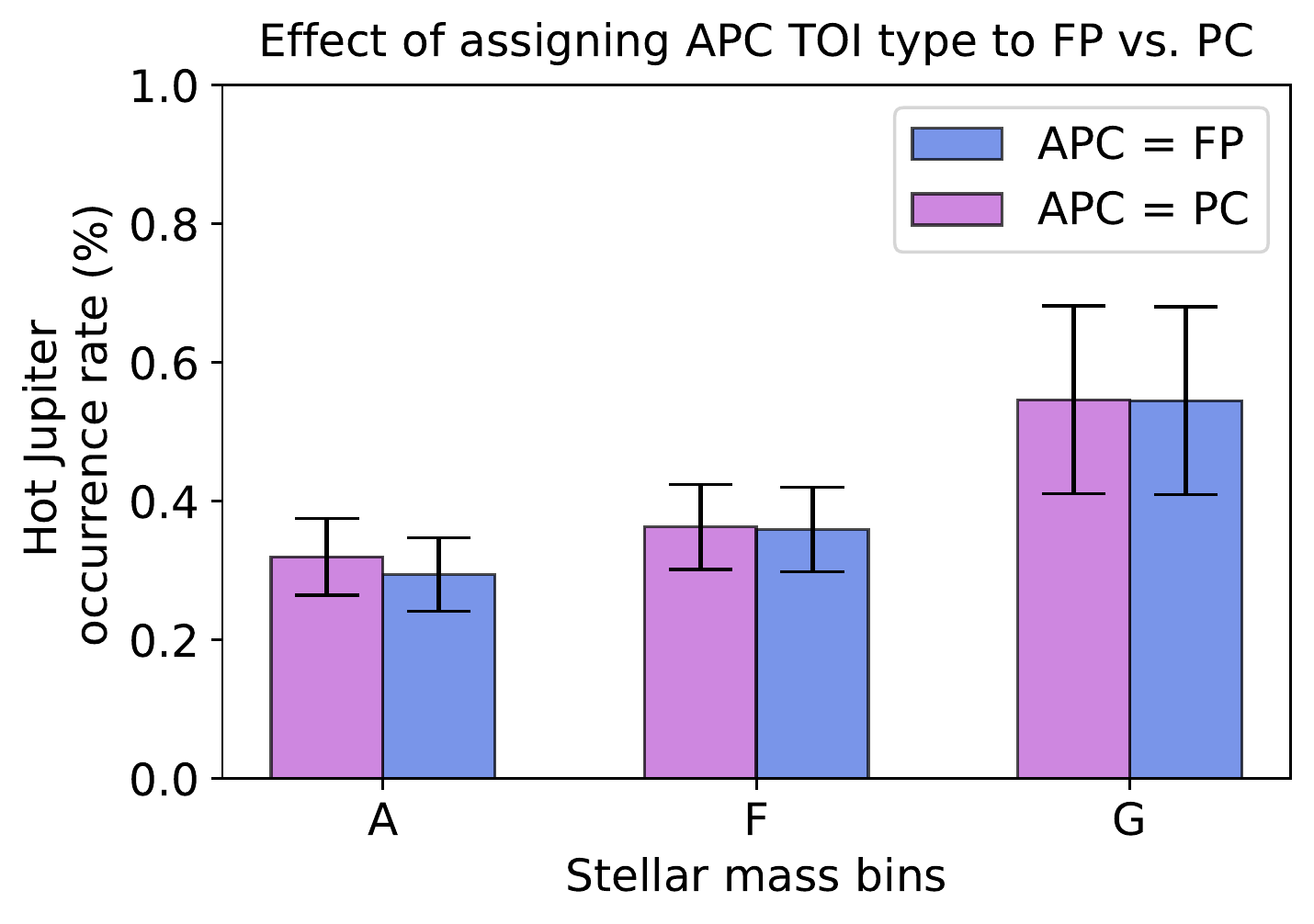}
        \caption{Hot Jupiter occurrence rates using our baseline assumption that ambiguous planet candidates (APCs) can be treated as false positives (FPs), compared to occurrence rates having treated APCs as PCs.  Treating APCs as PCs slightly increases our A star occurrence rate, but does not change any other results. There is no statistically significant difference between the results.}
    \end{figure}

        The false positive rate for A stars changed from $14 \pm 7\%$ to $5 \pm 5\%$, while F and G false positives rates stayed the same as there are no APCs in these stellar mass bins. The new A type occurrence rate is $0.32\pm0.06\%$, which is similar to our baseline value (Fig. \ref{fig6}). We conclude that this assumption did not affect the interpretation of our results.
    
	\subsection{Comparison with \textit{Kepler} Occurrence Rates}\label{sec4.4}
		\textit{Kepler} occurrence rate works have primarily focused on F, G, and K dwarf stars, allowing us to compare F and G occurrence rates between TESS and Kepler. While there may be differences due to the fact that \textit{Kepler} occurrence rate studies predominantly selected stellar samples based on simple log$g$ rather than isochrones cuts, meaning there may be mischaracterized subgiants included in \textit{Kepler} stellar samples, we believe this effect will be minimal given F and G star results were still well within $1\sigma$ after changing our stellar sample selection.
		
        Our average of F and G occurrence rates, $0.39 \pm 0.06\%$, is in strong agreement with the $0.43\pm0.05\%$ FGK hot Jupiter occurrence rate from \citet{fressin2013}. We also find our $0.55 \pm 0.14\%$ for G type stars has better than $1\sigma$ agreement with the $0.4\pm 0.1\%$ GK abundance by \citet{howard2012} and the more recent $0.57^{+0.14}_{-0.12}\%$ found by \citet{petigura2018}.
		
		\citet{kunimoto2020} determined exoplanet occurrence rates from \textit{Kepler} for planets up to $16 R_{\oplus}$ around FGK stars. Summing their occurrence rates for $R_{p} = 8 - 16 R_{\oplus}$ and $P = 0.78 - 12.5$ days as a close representation of hot Jupiters, \citet{kunimoto2020} found occurrence rates of $0.21 \pm 0.10\%$ and $0.52\pm 0.18\%$ for F and G type stars, respectively. While our F type occurrence rates are slightly ($1.3\sigma$) higher, which may be due to differences in the planet radius regime covered by our works, our G type occurrence rates are well within $1\sigma$ and both works indicate higher hot Jupiter abundance with lower stellar mass.
		
		Overall, we find strong agreement between TESS and \textit{Kepler} hot Jupiter occurrence rates.
		
		\subsection{Comparison with RV Occurrence Rates}\label{secRV}
		
		There is a well-known discrepancy in hot Jupiter occurrence rates between \textit{Kepler} and RV surveys, where RV studies typically find abundances of $\sim0.8 - 1.2\%$ \cite[e.g.][]{mayor2011, wright2012, Wittenmyer2020}, a factor of two higher than those from \textit{Kepler}. Different host star metallicities \cite[e.g.][]{wright2012, hinkel2014, Guo2017}, evolutionary stages \cite[e.g.][]{johnson2010}, and false positive rates \cite[e.g.][]{wang2015} between transit and RV surveys have been explored in literature, yet none have been able to fully account for the discrepancy.
		
		Another explanation involves the differences in stellar multiplicity within typical stellar samples \cite[e.g.][]{wang2014, Moe2021}. This proposed solution lies in the fact that RV surveys are significantly biased against binaries, especially close spectroscopic binaries, given that even a few measurements may be enough to remove stars with large-amplitude RV variability due to stellar-mass companions. Meanwhile, transit surveys do not make significant selection biases against binaries. The hot Jupiter abundance from RV surveys can thus be interpreted as the abundance around single stars, while transit surveys can be interpreted as the abundance around all stars. Because the presence of a stellar companion is expected to suppress the formation of close-in giants \cite[e.g.][]{wang2014}, the single-star (RV) hot Jupiter occurrence rate should be higher than the all-star (transit) occurrence rate.
		
		\citet{Moe2021} estimated that A and G hot Jupiter planet occurrence rates for single stars should be $2.1\pm0.3$ and $1.8\pm0.2$ times larger than those for all primaries in magnitude-limited surveys. When applying these corrections, we find TESS hot Jupiter occurrence rates of $0.62 \pm 0.32\%$ and $0.98 \pm 0.36\%$ for A and G stars, with our G-type occurrence rate well in line with results from RVs for Sun-like stars. In other words, the correction for stellar multiplicity alone appears capable of accounting for the discrepancy between RV and transit methods.
		
		RV surveys have also found that giant planet occurrence rates increase with stellar mass \cite[e.g.][]{Bowler2010, johnson2010, Gaidos2013}. Both our baseline AFG TESS results and FGK \textit{Kepler} results from \citet{kunimoto2020} show evidence for the opposite, where hot Jupiters are more common around lower mass stars. Even after multiplicity correction, we still do not find evidence for a positive correlation between hot Jupiter abundance and stellar mass. 
		
		The RV giant planet samples used to establish the trend include those in orbits out to several AU, and we do not expect that trends among long-period giants are necessarily the same for short-period giants. \citet{Bowler2010} found significantly higher abundances of giant planets within 3 AU around evolved A stars than Sun-like stars, but found a paucity of giants within 0.6 AU, implying unique planet formation or evolution processes are at play. A possible explanation is that the enlarged radii of evolved stars have engulfed or tidally disrupted short-period planets. \citet{Hamer2019} found that hot Jupiters could be destroyed by tides as early as the main sequence, which could explain why our A dwarf abundances are lower than those for G dwarfs. We do not believe that our hot Jupiter occurrence rates are in tension with trends in giant planets from RVs.
		
\section{Conclusion}\label{sec5}

We have presented hot Jupiter occurrence rates based on 97 hot Jupiters detected with an independent search of TESS Prime Mission light curves. Our stellar sample consists of 198,721 bright ($T < 10.5$ mag) AFG main sequence stars observed over the TESS Prime Mission. We find occurrence rates of  $0.29 \pm 0.05\%$ for A type stars, $0.36 \pm 0.06\%$ for F type stars, and $0.55 \pm 0.14\%$ for G type stars, with an overall weighed average across stellar types of $0.33 \pm 0.04\%$. This more than halves the uncertainties in TESS occurrence rates found by \citet{zhou2019} across all three stellar mass bins. In doing so, we find that hot Jupiter abundance decreases as stellar mass increases, consistent with trends seen for planets from Kepler \citep{mulders2015, kunimoto2020}. Our results, as supported with findings by \citet{Hamer2019}, could indicate that hot Jupiters are less common around larger stars due to increased tidal disruption.

Our F and G occurrence rates are in good agreement with FGK hot Jupiter abundances from \textit{Kepler} ($\sim0.4 - 0.6\%$), which are roughly half of the typical rates found by RV studies ($\sim0.8 - 1.2\%$) for Sun-like stars. This is despite the fact that the TESS stellar sample is more similar to typical RV targets (bright stars in the solar neighbourhood) than \textit{Kepler} stars. When using results from \citet{Moe2021} to correct for stellar multiplicity in our magnitude-limited stellar sample, we find G occurrence rates of $0.98 \pm 0.36\%$, well within $1\sigma$ of estimates from RV. Correction for stellar multiplicity is thus capable of fully accounting for the discrepancy between RV and transit methods.

There remains much to be learned from TESS. Here we utilize data only from the Prime Mission, but TESS has since observed for nearly two more years, including in new areas of the sky. The sample of hot Jupiter TOIs will continue to grow. Furthermore, while our focus was on stars brighter than 10.5 mag, TESS data may give significantly more statistical power when including TOIs around fainter stars. In particular, down to the QLP magnitude limit of 13.5 mag, the hot Jupiter sample is now more than 1500 TOIs \citep{Kunimoto2022}. Analyzing fainter stars will also allow us to explore hot Jupiter occurrence rates for K dwarfs, and potentially M dwarfs. The broad reach of TESS will in turn allow for further improvement of hot Jupiter demographics and the precision of exoplanet occurrence rates.

\section*{Acknowledgements}

We thank the referee for their constructive comments, which significantly improved the quality of the paper.

This paper utilizes data from the Quick Look Pipeline (QLP) at the TESS Science Office (TSO) at MIT. The TESS mission is funded by NASA’s Science Mission Directorate. 

This paper includes data collected with the TESS mission, obtained from the MAST data archive at the Space Telescope Science Institute (STScI). Funding for the TESS mission is provided by the NASA Explorer Program. STScI is operated by the Association of Universities for Research in Astronomy, Inc., under NASA contract NAS 5–26555. This research has made use of the Exoplanet Follow-up Observation Program website \citep{ExoFOP}, which is operated by the California Institute of Technology, under contract with the National Aeronautics and Space Administration under the Exoplanet Exploration Program. This work has made use of data from the European Space Agency (ESA) mission {\it Gaia} (\url{https://www.cosmos.esa.int/gaia}), processed by the {\it Gaia} Data Processing and Analysis Consortium (DPAC, \url{https://www.cosmos.esa.int/web/gaia/dpac/consortium}). Funding for the DPAC has been provided by national institutions, in particular the institutions participating in the {\it Gaia} Multilateral Agreement.

This research made use of \texttt{exoplanet} \citep{exoplanet:joss,
exoplanet:zenodo} and its dependencies \citep{exoplanet:agol20,
exoplanet:arviz, exoplanet:astropy13, exoplanet:astropy18, exoplanet:kipping13,
exoplanet:luger18, exoplanet:pymc3, exoplanet:theano}.

\section*{Data Availability}

The analyzed hot Jupiters and their properties are available in the article and in its online supplementary material. QLP light curves and the CTL catalog are available on MAST.



\bibliographystyle{mnras}
\bibliography{example} 





\bsp	
\label{lastpage}
\end{document}